\documentclass[aps,twocolumn,amsmath,amssymb,prl,showpacs]{revtex4-1}
\usepackage{graphicx}

\usepackage{bm}

\def\beq{\begin{equation}}
\def\eeq{\end{equation}}

\begin{document}

\title{Unruh Effect in non local field theories}

\author{Nirmalya Kajuri}
\email{nirmalya@physics.iitm.ac.in}
\affiliation{Department of Physics, Indian Institute of Technology Madras, Chennai 600036}

\bibliographystyle{apsrev4-1}

\begin{abstract}

We study Unruh effect for a wide class of non local theories. Using the approach of Bogoliubov coefficents we show that Unruh effect remains entirely unmodified in these theories. However, for those non local theories which incorporate a minimal length, the approach using Unruh-DeWitt detectors  predicts a modification in Unruh effect.  Previous work shows that this modification may even be drastic. This appears contradictory, with two apparently equivalent methods giving different results. We investigate the origin of the contradiction and show that for these theories, the two methods are indeed inequivalent. We argue that the detector method incorporates an assumption of local interaction that does not hold for non local theories and therefore the method of Bogoliubov coefficents is preferable. 

\end{abstract}
\maketitle
\paragraph{Introduction.}Interest in non local theories dates back to the early days of quantum field theory (See \cite{Efimov:1977bpa} for an early review). Recently there has been a resurgence of interest in these theories particularly in the context of non local gravity theories \cite{Tomboulis:1997gg,Biswas:2010zk,Moffat:2010bh, Biswas:2011ar,Modesto:2011kw, Modesto:2014lga,Frolov:2015bia,Frolov:2015bta} and cosmological models\cite{Deser:2007jk,Barnaby:2007ve,Barnaby:2008fk,Barnaby:2010kx, Deser:2013uya}. For a review of these theories we refer the reader to \cite{Tomboulis:2015gfa}. In this paper we investigate Unruh effect in a class of non local scalar field theories. 

Previously, a particular non local theory had been considered by  Nicolini and Rinaldi in \cite{Nicolini:2009dr} considered a particular minimal length model which incorporated a minimal length.  The particular model Nicolini and Rinaldi considered was defined by the  propagator: 
\beq
\label{nl} G_l(p^2) = \frac{e^{-l^2p^2}}{p^2}
\eeq
For this propagator Nicolini and Rinaldi had calculated the response rate of an accelerating Unruh-DeWitt detector \cite{Unruh:1976db,DeWitt:2003pm}and showed a drastic decrease in the rate, even for very small $l$

In general, it is quite evident that for \eqref{nl} and similar propagators,  Unruh effect as computed by the approach of a Unruh-DeWitt detector must give a result that is a  function of the minimal length.  The response rate of an Unruh-DeWitt detector is calculated from the two point function of the field, and if the two point function depends on $l$ so will the response rate. It is clear that the result obtained by this approach must differ from the result for the usual Klein Gordon field. 

A  different approach to deriving the Unruh effect involves calculating the Boguliubov coefficients between the Minkowski and Rindler space solutions \cite{Unruh:1976db}. To see how this method can be applied to the theory given by \eqref{nl} we note that it  corresponds to the infinite order differential equation\footnote{The operator $e^{-l^2\Box}$ is not well defined in a spacetime with Minkowski signature. However $e^{-l^2\Box^{2n}}$ is a valid operator for any positive $n$}: 
\beq
\label{nle} \Box \left(e^{-l^2\Box} \phi\right) =0
\eeq

The set of solutions for equations like \eqref{nle} can be deduced for both Minkowski and Rindler spacetimes. This enables us to calculate the required Boguliubov cooefficients. As we will see, the Boguliubov coefficients are exactly the same as in the case of a massless Klein Gordon field.  This approach then predicts no modification of the usual Unruh effect for non local field theories. 

We will prove this result for a wider class of non local field theories, which may or may not incorporate a single characteristic length scale. But for those which do, clearly the results of this approach cannot match the ones coming from the Unruh-DeWitt detector calculation. This seems to be a paradoxical result, with two different methods of deriving Unruh effect giving entirely different predictions. While the two approaches are not expected to match for co-ordinates where positive norm solutions differ from positive frequency solutions \cite{Sriramkumar:1999nw, Sriramkumar:2016nmn}, such is not the case here. We will argue that the origin of this discrepancy likely lies in modelling the detector by a local observable in a non local theory. 

In the next section, we introduce a class of non local massless scalar field theories and prove that the Unruh Effect remains unchanged in these theories.  In the third section we address the tension between the two approaches of deriving Unruh effect and offer a possible resolution to the paradox. 

\paragraph{Non local field theories and Unruh effect}

We introduce the class of non local theories which we will consider. These will be given by the following homogeneous, infinite order differential equation: 
\beq
\label{iokg}\Box(F(\Box) \phi) = 0
\eeq

where the assumption of $F$ will be that it is an analytic function that is everywhere non zero.

The propagator of such a theory is given by: 
\beq
\label{pro} G(p^2) = \frac{F(-p^2)^{-1}}{p^2}
\eeq

Clearly the theory governed by  \eqref{nle} falls in this class, as does any theory of the form  $\Box \left(e^{-l^2\Box^{2n}}\right) \phi =0$

Now we would like to find the solutions for any such theory. In this we are helped by the following result\cite{Barnaby:2007ve}: the number of independent solutions of an infinite order differential equation is equal to the number of poles in its propagator, just as in usual field theories. Particularly, for a propagator of the form \eqref{pro}, the number of independent solutions will be two. 

 Now note that a solution to  $\Box\phi =0$ would also be a solution to \eqref{iokg}. But we saw above that \eqref{iokg} has only two solutions. it follows that the complete set of solutions to \eqref{iokg} are just the solutions of  $\Box\phi =0$. We call this fact result I.

The result above can be generalized to different metrics\cite{Barnaby:2007ve}. That is, if we write the counterpart of \eqref{pro} for a  spacetime with metric components  $g_{\mu\nu}$

\beq
\label{iokgc}\Box_g(F(\Box_g) \phi) = 0
\eeq

where $\Box_g = g^{\mu\nu} \nabla_{\mu}\nabla_{\nu} $, the covariant D'Alembartian.

Just as in the case of Minkowski metric, the complete set of solutions to this equation will be given by the solutions to $\Box_g\phi =0$.  In particular, this will be the case in Rindler space. 

With these results, the proof of robustness of Unruh effect in the Boguliubov coefficient method is obvious.  As the full set of solutions of theories given by \eqref{iokg}  matches the solutions of massless Klein Gordon equation in both Minkowski and Rindler spaces, it follows that the Bogoliubov coefficients also match. Consequently there is no modification to Unruh effect for these theories.

\paragraph{ A contradiction?}

Clearly, there is a  tension between the results obtained through the two different approaches which seems contradictory at first sight. The response rate of an Unruh-DeWitt detector must show a modification while we have proved above that the Bogoliubov coefficients remain unchanged. 

Mathematically, the origin of this tension is easy to see. The equivalence of the two approaches depends on the following relation between the two point function $G(x)$ and the set of solutions $u_i(x)$:
\beq
\label{equ}G(x-x') = \sum_i u_i(x)u_i^{*}(x')
\eeq
This result does not hold for infinite order differential equations. Generally, given a propagator for a finite order differential equation (let's take $1/p^2$ as a concrete example)  multiplying to it an analytic function $F(-p^2)$ which is nowhere zero gives us the propagator for an infinite order differential equation of the form \eqref{iokg}. However such a multiplication has no effect on the set of solutions of the original equation; the new equation  \eqref{iokg} will have the exact same set of solutions. Thus it is only the left hand side of \ref{equ} that gets modified. 

While this shows us why the two approaches cannot agree, it raises the further question: is one to be preferred over the other? We would like to argue that in this case, the method of Unruh-DeWitt detector is not physically meaningful.
This is because in the other method, the detector is modelled as a point-like system which only interacts with the field at the point where it is located. However a non local field as we have considered here would always interact non-locally with any detector. In other words, a point-like detector would in general be affected by the field value at points away from it's location. Consequently, modelling the detector as a local observable is inappropriate for a non local theory.

In other words, the mistake lies in describing the detector-field interaction by the Green function of the \textit{free} theory. Interactions have to be taken into account. As shown in \cite{Tomboulis:2015gfa}, the theory can then be reformulated in terms of the standard field propagator and non local interaction vertices. Taking these into account, one can calculate Unruh effect in non local theories correctly by considering detectors of finite size much larger than $l$. The interaction vertices would be smeared over this length scale. The coarse-grained theory would appear as local and therefore such a computation should agree with the results obtained from the Boguliubov coefficients.

To conclude, we have shown that Unruh effect remains unmodified for a wide class of non local theories. This result seemed to contradict the expectation for those non local theories which incorporate dimensionful parameters (such as a minimal length). We pinpointed the origin of this contradiction and pointed out that  modelling detectors as local observables is not physically meaningful in non local theories.																																																																																					

\paragraph{ Acknowledgements} It is a pleasure to thank Dawood Kothawala, Alok Laddha and Gaurav Narain for helpful discussions.We would also like to thank the anonymous referee for their suggestions. 
\bibliography{unruh}
\end{document}